\documentclass[sigconf]{acmart}
\usepackage{framed}
\usepackage{delarray}
\usepackage{subfigure}
\AtBeginEnvironment{quote}{\par\singlespacing\small}

\usepackage{draftwatermark}
\SetWatermarkText{DRAFT}
\SetWatermarkScale{0}

\AtBeginDocument{%
  }

\copyrightyear{2026}
\acmYear{2026}
\setcopyright{cc}
\setcctype{by}
\acmConference[DIS '26]{Designing Interactive Systems Conference}{June 13--17, 2026}{Singapore, Singapore}
\acmBooktitle{Designing Interactive Systems Conference (DIS '26), June 13--17, 2026, Singapore, Singapore}
\acmDOI{10.1145/3800645.3812954}
\acmISBN{979-8-4007-2563-0/2026/06}

\begin{document}

 

\title[How Lived Experiences are Entangled with AI Predictions in Menstrual Cycle Tracking Apps]{"It became a self-fulfilling prophecy”: How Lived Experiences are Entangled with AI Predictions in Menstrual Cycle Tracking Apps}

\author{Wendy Zhou}
\orcid{0009-0008-6107-7198}
\affiliation{
 \institution{Department of Computer Science and Engineering, Chalmers University of Technology \\ University of Gothenburg}
 \streetaddress{}
 \city{Gothenburg}
 \country{Sweden}
 \postcode{SE-41296}
 }
 \email{zwendy@chalmers.se}

\author{Pelin Karaturhan}
\orcid{0000-0002-3841-3463}
\affiliation{
\institution{IT University of Copenhagen}
\streetaddress{Rued Langgaards Vej 7}
\city{Copenhagen}
\country{Denmark}
\postcode{DK-2300}
 }
\email{peka@itu.dk}

\author{Alexandra Weilenmann}
\orcid{0000-0002-3989-4588}
\affiliation{
\institution{Department of Applied Information Technology, University of Gothenburg}
\city{Gothenburg}
 \country{Sweden}
\postcode{405 30}
 }
\email{alexandra.weilenmann@gu.se}
 
\author{Jichen Zhu}
\orcid{0000-0001-6740-4550}
\affiliation{
\institution{IT University of Copenhagen}
\streetaddress{Rued Langgaards Vej 7}
\city{Copenhagen}
 \country{Denmark}
\postcode{DK-2300}
 }
\email{jicz@itu.dk}

\renewcommand{\shortauthors}{Zhou, Karaturhan, Weilenmann, and Zhu}

\begin{abstract}

In menstrual cycle tracking apps (MCTAs), AI-based predictions and insights have become increasingly popular. These features enable users to receive personalized information about their bodies and mental states. However, there is currently little research on how these predictive AI features and explanations affect users' lived experiences. This paper examines human-AI entanglement in MCTAs through 14 semi-structured user interviews and a group autoethnography. These methods uncover the processes leading to this phenomenon. Our results reveal that: (1) users understand their lived experiences in light of AI predictions, although these predictions can be faulty due to imperfect logging practices, (2) the user interface features and AI explanations do not support awareness or critical engagement with this entanglement and meaning-making, and (3) non-normative MCTA users report a sense of isolation in this entangled interaction. Based on our findings, we propose design implications for predictive AI features and explanations.


\end{abstract}


\maketitle


\section{Introduction}

\begin{figure*}[t]
    \centering
    \subfigure{\includegraphics[width=0.24\textwidth]{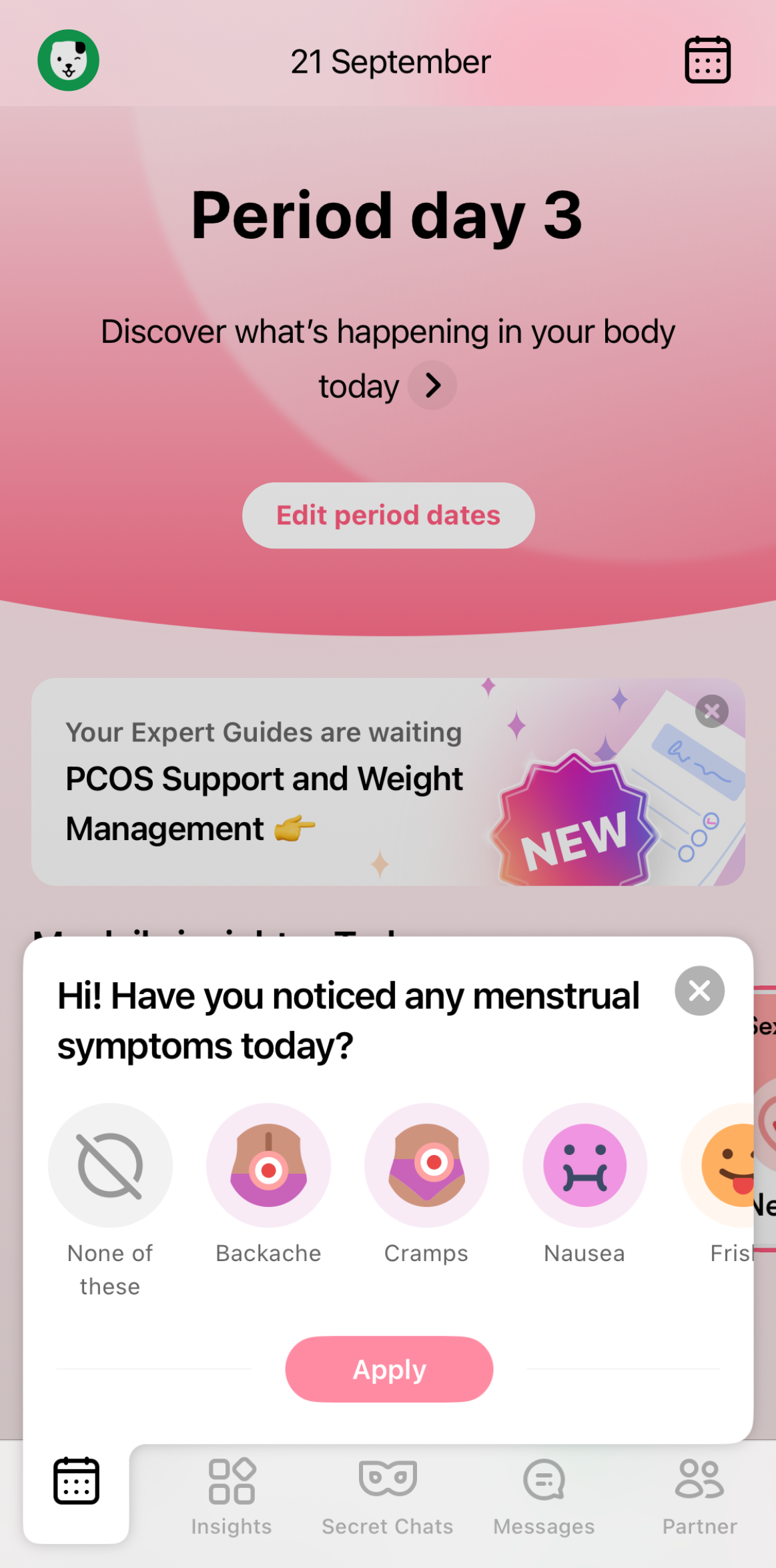}} 
    \subfigure{\includegraphics[width=0.24\textwidth]{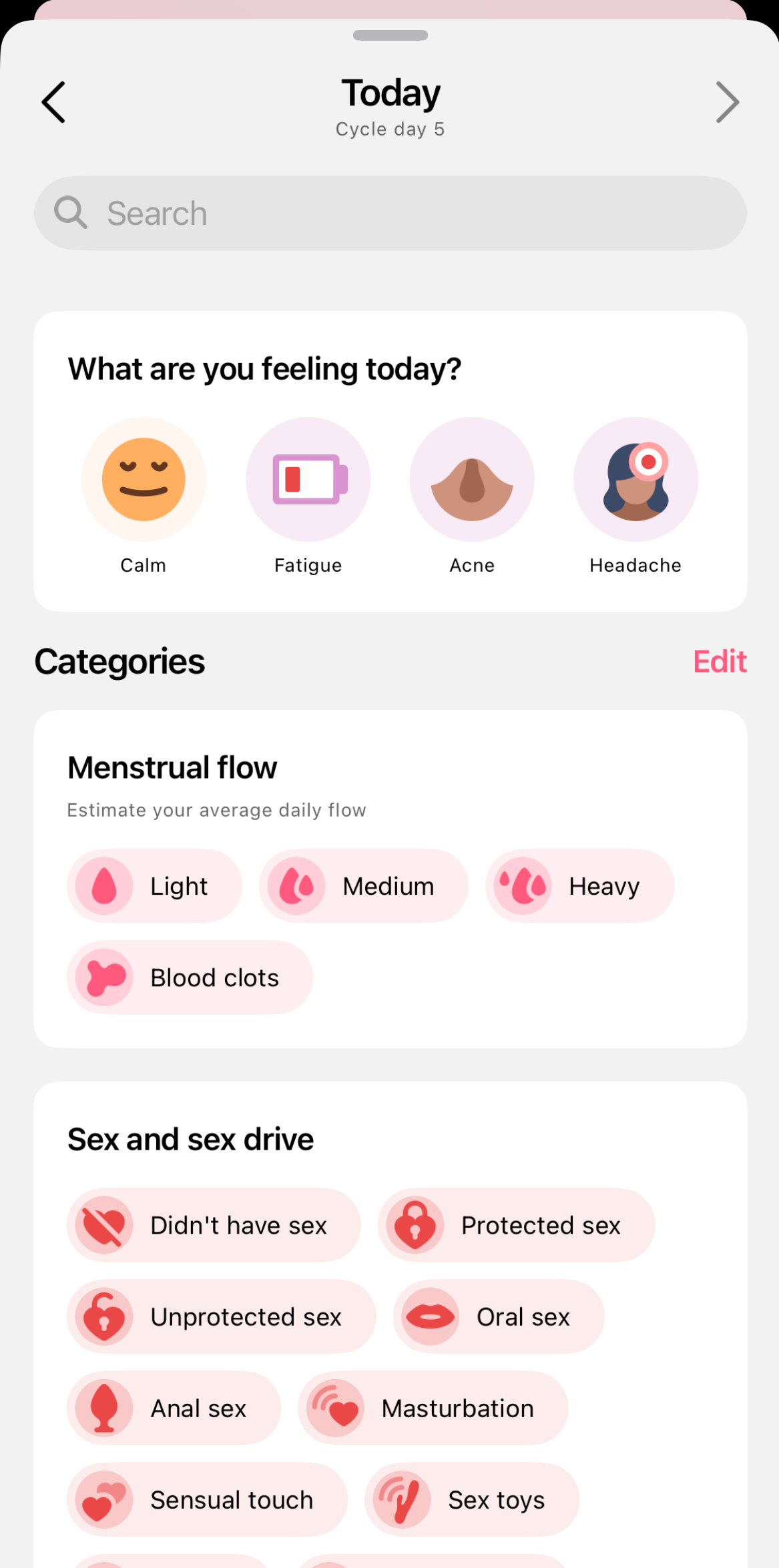}} 
    \subfigure{\includegraphics[width=0.24\textwidth]{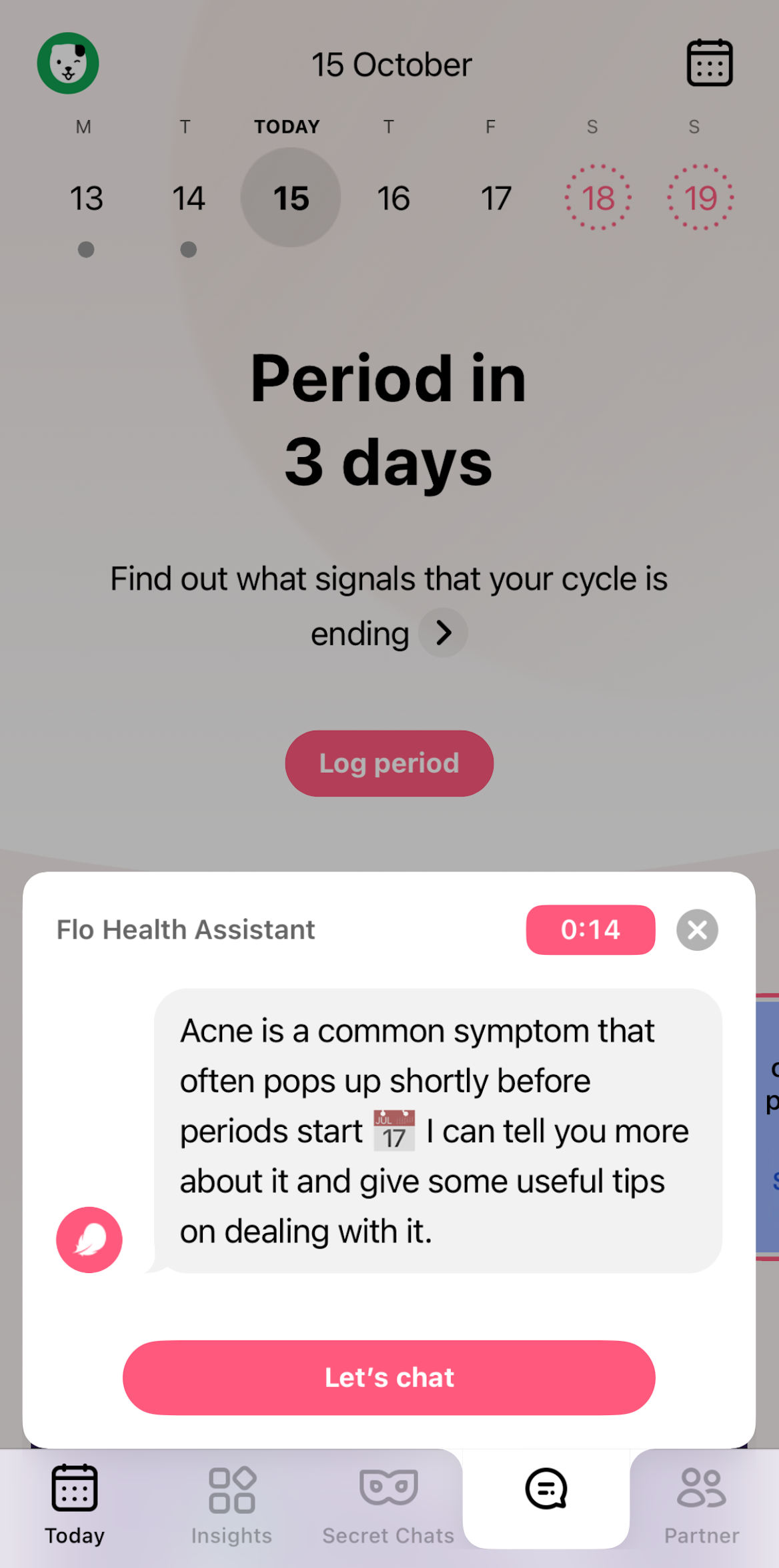}}
    \subfigure{\includegraphics[width=0.24\textwidth]{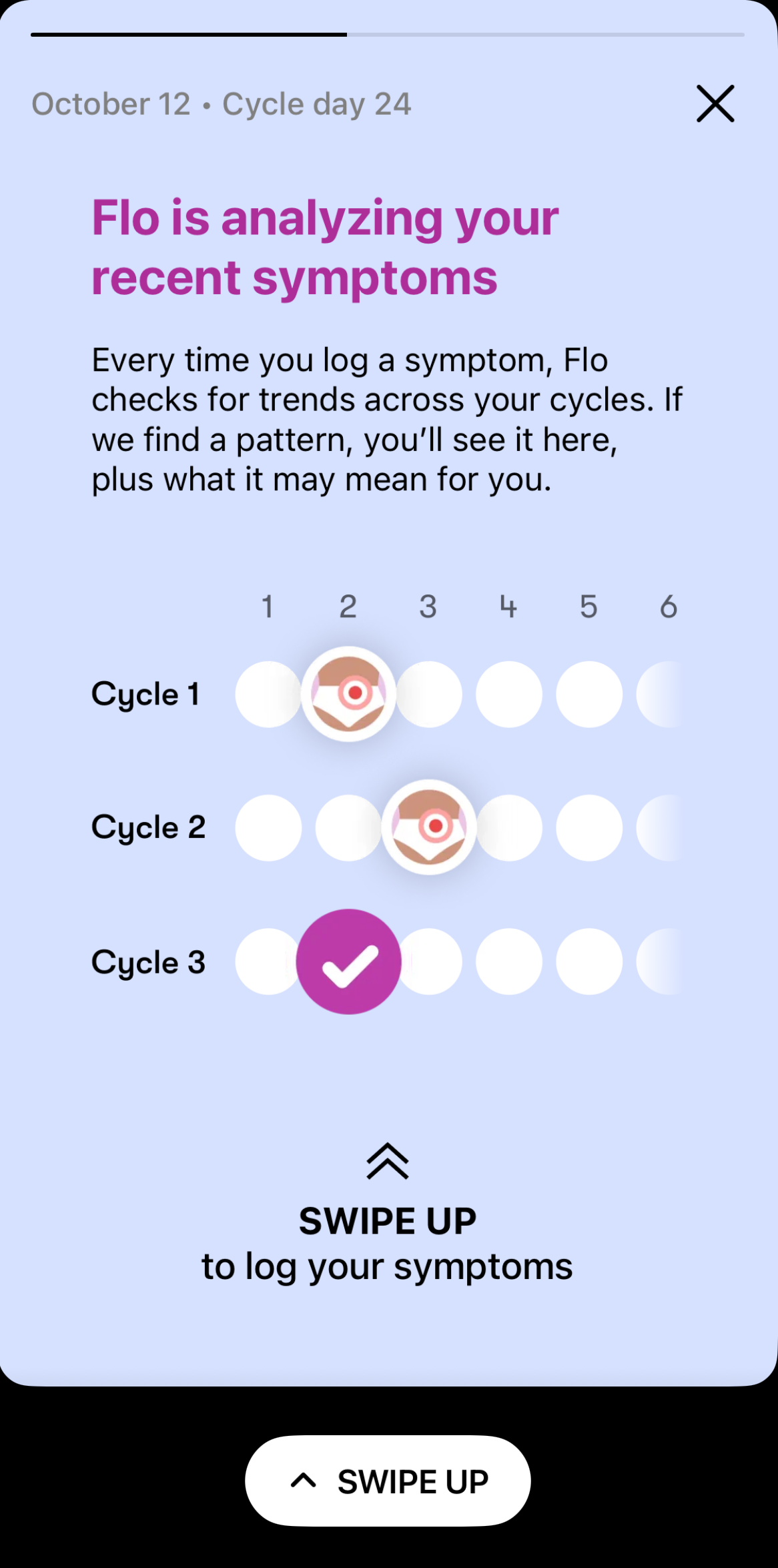}}
    \Description[Four screenshots from the Flo menstrual cycle tracking app illustrating its symptom logging, daily tracking, AI chatbot, and prediction explanation features.]{Four screenshots from the Flo menstrual cycle tracking app. (a) A list of suggested symptoms and moods to log for the day, including items such as Backache, Cramps, and Nausea. (b) The main logging interface with categories for recording daily experiences and menstrual flow. (c) A chatbot interface displaying AI-driven predictions and personalized recommendations. (d) An AI explanation screen describing how the app uses personal data to generate cycle predictions.}
    \caption {Four screenshots from the {\em Flo} menstrual cycle tracking app showing: (a) a list of suggested symptoms and moods to log for the day including items like 'Backache,' 'Cramps,' and 'Nausea'; (b) the main logging interface with categories for logging daily experiences and menstrual flow; (c) a chatbot interface displaying AI-driven predictions and personalized recommendations; and (d) an AI explanation screen describing how the app uses data to generate predictions.}
    \label{fig:AI_Features}
\end{figure*}

Fertility and menstrual cycle tracking apps (MCTA), such as commercially available {\em Flo}, {\em Clue}, and {\em Stardust}, have become increasingly prevalent. In these well-being technologies, AI-based features allow users to receive personalized predictions, insights, and recommendations on various aspects of the menstrual cycle and fertility (See Figure \ref{fig:AI_Features} for screenshots from the {\em Flo} MCTA). These predictive AI features focus on pre-menstrual (PMS) and menstrual symptoms, such as moods and mental states (e.g., depression, anxiety, and brain fog), and physical symptoms  (e.g., cramps, bloating, cervical fluids, and menstruation), based on users' self-tracking.

Although MCTAs can be a valuable source for personalized health and well-being management, these predictive AI features have several critical shortcomings, especially a lack of transparency into how predictions are generated and limited mechanisms for users to correct them \cite{mirzaliyeva2024ai}. Research has shown that the complexity and black-box nature of AI make it difficult for non-technical users to understand its decision-making and to develop an appropriate level of trust and reliance \cite{jacovi_formalizing_2021}. In mobile health and well-being applications, AI is still far from 100\% accurate in many tasks \citep{dratsch2023automation,zhu2025centers}, including predicting menstruation cycles and related symptoms. MCTAs' widespread adoption in the general population means many users may face challenges using these AI features appropriately and responsibly. 

A promising approach to increasing the transparency of AI is eXplainable AI (XAI) ~\cite{gunning2019darpa}. The rapid growth in XAI research has led to a wide range of ML techniques that aim to enable human users to understand, appropriately trust, and produce more explainable models and explanation interfaces \cite{dwivedi2023explainable,haque2023explainable}. Despite the technical advancement, however, XAI research has limited success in real-world applications such as MCTAs and other well-being technologies due to the low usability and efficacy of the provided explanations ~\cite{Abdul2018TrendsSystems, Doshi-Velez2017AInterpretability,miller2019explanation,Zhu2018,nguyen2024human,haque2023explainable}. More alarmingly, recent studies have found when explanations for AI features are not well designed to meet user needs, they may lead to users' over-reliance on these technologies~\cite{lai2019human,papenmeier2019model}, causing potential harm. Therefore, a rich understanding of users' lived experiences with these technologies is the foundation for formulating user needs and designing responsible AI features and explanations. For the rest of the paper, we use "AI explanations" broadly to refer to any information, specifically designed for end users, about how AI features work in MCTAs and why certain predictions were made.

Recent research inquiries into understanding the lived experiences of users with self-tracking technologies and MCTAs in HCI field emphasize the uniqueness of the intersection between the bodily experience and interacting with the external information about the same body \cite{homewood_removal_2020, cosley_introduction_2017}. This research emphasizes that in the context of self-tracking, understanding the technology is not only limited to its performance and usability, but also its impact on subjective experiences of its users \cite{zhu2025centers,homewood_removal_2020,homewood_putting_2020,reime2023walking}. This understanding aligns with Entanglement HCI, a research perspective committed to the notion that humans and things are "ontologically inseparable from the start" \cite{introna2013towards}, and prioritizing phenomenological understandings of this entanglement \cite{frauenberger2019entanglement}. Within MCTA research, relatively little work has focused on predictive AI features for mood changes and other PMS symptoms. Compared with menstrual periods, these symptoms are less objectively identifiable, and only the users themselves know the "ground truth". The subjective nature of moods and PMS symptoms opens up space for AI predictions to affect the users' felt experiences; for instance, users may trust AI predictions more than their own bodily sensations, which is a form of over-reliance to technology \cite{obrien_periods_2025}. Consequently, these predictive AI features should be studied not only in terms of their accuracy, but also in terms of their impact on users' subjective reality. This understanding of what we call the \textit{human-AI entanglement} is critical for designing ethical and responsible AI features. 



This work extends existing literature by investigating the entanglement between users and AI in the context of MCTAs, particularly for mood and symptom predictions. More specifically, this paper examines how predictive AI features and AI explanations shape users' lived experiences. In our study, the concept of entanglement is used to guide and inform our analysis, surfacing the ways human-AI interaction co-construct users' lived experience in the context of MCTAs. We address the following {\bf Research Question (RQ)}: {\em What kinds of understandings and experiences emerge as users engage with predictive AI features and AI explanations around moods and PMS symptoms in MCTAs?} 

To answer our RQ, we employ a qualitative approach by combining two studies. First, we conducted 14 semi-structured user interviews with regular MCTA users in northern and central Europe. Our analysis reveals that users' entanglement with predictive AI features in MCTAs co-constructs their lived experiences, and that they have intuitive understandings of AI features, despite their limited transparency. In addition, they seek validation and doubt their own experiences when predictions do not match what they feel in their bodies. To examine how the process of entanglement with predictive AI features happens through everyday use of MCTAs, we conducted a second study, a group autoethnography, where we, the authors, with diverse conditions including regular cycles, irregular cycles with Polycystic Ovarian Syndrome (PCOS), perimenopause, and menopause, tracked moods and symptoms in MCTAs over eight weeks. The autoethnography reveals further insights, including identifying when and how AI predictions and interface design influence and shape lived experiences, as well as how users with non-normative cycles experience marginalization. 

This paper makes the following contributions: (1) We reveal the complex human-AI entanglement processes through which users' lived experiences and self-knowledge are not only affected by, but also co-constituted with MCTAs through logging practices, interface designs, predictive AI features, and AI explanations. (2) We demonstrate that users lack resources (such as AI explanations and feedback mechanisms) to establish awareness and critical engagement with this human-AI entanglement phenomena. (3) We show the isolating experiences of non-normative users. (4) We propose alternative approaches to designing predictive AI features that support users' awareness and critical engagement with their entanglements with AI, while creating more inclusive designs for diverse users.

The paper proceeds as follows: First, we review related work. Then, we present Study 1 (user interviews) with its methods and results, followed by Study 2 (group autoethnography) with its methods and results, and conclude with a discussion that synthesizes the findings of both studies and discusses the design implications for predictive AI features in intimate health contexts.

\section{Related work}
In this section, we summarize related work on well-being technologies, including MCTAs, predictive AI features, AI explanation design, and entanglement approaches in HCI.

\subsection{AI Predictions and Explanations in Well-being Technologies}
Well-being technologies are tools, devices and platforms developed to help their users manage or improve their quality of life in a variety of domains, such as sleep quality \cite{son_balancing_2023}, physical activity \cite{keppel_situated_2025,zhu2021personalization}, mental health \cite{lee_designing_2025}, menstrual cycles and fertility \cite{park_ambivalences_2023, homewood_ovum_2019, homewood_putting_2020, campo_woytuk_touching_2020}, and chronic health conditions \cite{keys_rethinking_2025}. AI-based features are increasingly common in these technologies to support functionalities for data collection, classification, personalization, and prediction \cite{zhu2025centers}. Use of AI features can increase convenience and accuracy \cite{yan2022emoglass}, reduce cognitive load \cite{choe_semi-automated_2017}, and support self-awareness \cite{shin_toward_2025}.

However, these predictive insights can also lead to negative user experiences. The existing AI models in self-tracking well-being technologies are not 100\% accurate. Research has shown that even for relatively accurate models, their black-box nature may lead to users' inappropriate trust levels \cite{joseph_the_2025, figueiredo_powered_2024}. These models often simplify complex experiences \cite{wieczorek_ethics_2023}, resulting in mismatch with users' lived experiences \cite{nolasco_ai_2025} and overriding them \cite{zhu2025centers} 
to users' frustration \cite{joseph_the_2025}. These risks are also applicable in the context of MCTAs, especially when they are used for tracking subjective and intimate experiences such as moods and mental health symptoms. 

However, relatively little is known about how users engage with predictive AI for mental states in the context of well-being technology. This gap is essential, since unlike physiological states such as menstruation, which have observable physical markers, psychological states such as anxiety rely on subjective self-assessment, making them particularly dependent on self-interpretation \cite{sanches_hci_2019}. In other words, only the user themselves know the "ground truth" of how they feel. A user's mental states are therefore particularly susceptible to the influence of AI predictions, raising concerns about over-reliance on AI. This research aims to surface how human-AI entanglement operate in this under-explored domain.

A promising direction to improving the transparency of AI is eXplainable AI (XAI)~\cite{gunning2019darpa}. The rapid growth in XAI research has led to a wide range of ML techniques that aim to enable human users to understand, appropriately trust, and produce more explainable models \cite{dwivedi2023explainable,haque2023explainable}. However, most current XAI work focuses on technical development rather than producing adequate {\em explanations} for user needs \cite{Chromik2021, liao2021human}. As a result, most {\em explanations} produced by XAI still lack usability, practical interpretability, and efficacy for real users~\cite{Abdul2018TrendsSystems, Doshi-Velez2017AInterpretability,miller2019explanation,Zhu2018,nguyen2024human,meyer2024slide, haque2023explainable}. While a growing number of HCI research has investigated the design of AI explanations (e.g., \cite{Chromik2021,liao2021human,meyer_slide_2024,villareale2024can,cirqueira2020scenario,xu_explainable_2019}), the study of AI explanations in commercial well-being applications is underexplored in the HCI field. Two recent studies are particularly relevant. Su et al. analyzed the app descriptions and user reviews of 40 AI-supported mobile health applications and identified a lack of explanations to accurately describe AI features, leading to a mismatch between user expectations and the AI features \cite{figueiredo_powered_2024}. In the domain of MCTAs, Tylstedt et al. \cite{tylstedt_reimagining_2023} employed the method of critical app-walkthroughs by examining the AI explanations existing in various consumer applications, and found that the designs frame menstrual cycles as trackable, medical, and predictable, which the authors criticize for their techno-medical definition and lack of embodied perspective. Building on these existing studies, this paper provides a new design approach that acknowledges and works with human-AI entanglement, informed by our empirical findings on how MCTA users and predictive AI features are entangled. 

\subsection{Menstrual Cycle Tracking Apps (MCTA) and Lived Experiences}
MCTAs have received much attention in the HCI and design research community \cite{homewood_reframing_2018, GrimmeMyData2024, HudigDataSharing2025, fox_vivewell_2019, reime2023walking}. A seminal review paper by Epstein et al. \cite{epstein_examining_2017}  identified five reasons users track menstrual cycles: bodily awareness, cycle awareness, preparation, conceiving, and healthcare communications. Further studies have shown that MCTAs can increase users' understanding of their experiences \cite{punzi_mapping_2025, GrimmeMyData2024} and provide a sense of control \cite{obrien_periods_2025}.

However, the underlying assumptions, design decisions, and algorithmic approaches shaping these systems have received numerous criticisms. These apps and technologies are suggested to contribute to data extraction and menstrual surveillance \cite{gilman_periods_2021} and have become instruments of power and control \cite{punzi_mapping_2025}. Further, researchers have shown that MCTAs and their algorithms are designed with normative assumptions of end-users regarding to their cycle regularity, fertility, life stages, heterosexuality, ethnicity, religion, sexual-orientation, (dis)ability and gender, and can be exclusionary towards users different cycles, identities and bodies  \cite{sambain2025, reime2023walking, lupton2016quantifiedself, punzi_mapping_2025, ibrahim_islamically_2024, epstein_examining_2017}. Further, Fox et al. argued that by extracting intimate data and providing specific predictions, these apps reinforce transactional or instrumentalized perceptions of menstruating bodies \cite{fox_multiples_2020}. Through feminist participatory engagements to unfold lived experiences of menstrual tracking, they call for enacting sense-making, reflecting uncertainty, and foregrounding and unsettling norms in menstrual tracking technologies.

The limitations inherent in tracking inevitably affect the AI predictions users receive. This phenomenon calls for careful attention --- as previous research has argued that subjective experiences such as moods are affected by users' engagement with technology \cite{leahuSubjective}. In fact, O'Brien and Garcia Iriarte identified this co-construction in their qualitative study. Through semi-structured interviews, they identified a reliance on validating emotional states, which can escalate into an over-reliance, leading to inaccurate cycle predictions attributed to the body's faultiness rather than the app's \cite{obrien_periods_2025}. While their analysis sheds light on an essential phenomenon in users' interactions with MCTAs, they qualify this over-reliance in the context of inaccurate periods. By contrast, this paper aims to unfold the co-construction of lived experiences in AI-based mood and symptom prediction.

To address the limitations of menstrual cycle tracking technologies, the HCI and Design communities have adopted various epistemological approaches in alignment with Entanglement HCI \cite{frauenberger2019entanglement}, such as posthumanism \cite{zhu2025centers, homewood_removal_2020} and phenomenology \cite{homewood_putting_2020}, and have leveraged first-person perspectives \cite{reime2023walking, homewood_removal_2020, zhu2025centers}. For example, Homewood and Vallgårda \cite{homewood_putting_2020} follow a set of "phenomenological commitments" including that self-knowledge is co-constructed, that bodies are in a constant state of change, and that the self-knowledge of bodies is interconnected with that of others, to explore alternative designs that support subjective experiences. Recently, Zhu et al. critically examined the assumptions underlying machine learning models for three digital health technologies, including a menstrual-tracking app, using posthumanism as a lens. Through asking what is centered and what or who is left at the margins, they propose new ways of modeling humans in AI-supported well-being technologies \cite{zhu2025centers}. For menstrual tracking apps, this includes, for instance, designing systems that support self-exploration to recognize external factors (e.g., stress, illness, medication) that influence symptoms. They conclude by calling for machine learning algorithms that emphasize the entanglement between the technologies and people.

These studies call for further research into entanglement through the lens of design, emphasizing the importance of understanding how AI-supported MCTAs and human experience co-shape one another. We contribute to the emerging field of critical engagements, focusing on the lived experiences of MCTAs. We do this by conducting in-depth, semi-structured interviews and first-person accounts, with an emphasis on AI-based predictions and explanations, and on subjective experiences of moods and symptoms.

\section{Study 1: Interview Study}

\subsection{Methods}
We conducted an interview study with users who have been or are currently active MCTAs users to gain insight into the understandings and experiences that emerge from engagement with AI predictions and explanations. Semi-structured interviews were selected for their open-ended nature, which enables a deeper understanding of participants' experiences and allows for follow-up questioning. The participant sample size was based on similar qualitative interview studies, which reported a mean sample size of 14-16 participants \cite{cainesample2016}. A reflexive thematic analysis of the interviews (see Section \ref{interview_analysis}) was conducted to gain rich insights into the lived experiences of our participants with AI features and explanations.

\subsubsection{Participants}
\begin{table*}[t]
    \centering
    \begin{tabular}{>{\centering\arraybackslash}p{0.07\linewidth}>{\centering\arraybackslash}p{0.05\linewidth}>{\centering\arraybackslash}p{0.4\linewidth}>{\centering\arraybackslash}p{0.2\linewidth}>{\raggedright\arraybackslash}p{0.1\linewidth}>{\raggedright\arraybackslash}p{0.05\linewidth}}\toprule
         Participant& Age & Education& Apps Used&Period Type& Current User\\\midrule
         P1& 27&  Information Systems and Human-Centered AI& Clue  & Regular& Yes\\
         P2& 27&  Cognitive Science and Interaction Design& Stardust, Hormona, Maya & Irregular& Yes\\
         P3& 25&  Visual Arts& Clue & Regular& Yes\\
         P4& 31&  Geography and Cartography, Applied Climate Strategy& Flo, Natural Cycles & Regular& Yes\\
         P5& 25&  Economics and Human-Centered AI& Menskoll & Regular& Yes\\
         P6& 26&  Political Science& Apple Health & Irregular& Yes\\
         P7& 28&  Medicine (Doctor)& Huawei Health & Regular& Yes\\
         P8& 34&  Subject Teacher& Clue, Natural Cycles & Regular& Yes\\
         P9& 28&  Digital Management& Flo, Natural Cycles & Irregular& Yes\\
         P10& 28&  Cognitive Science and Interaction Design& Maya, Stardust, Flo, Menskoll& Regular& Yes\\
         P11& 30&  -& Clue, Flo & Irregular & Yes\\
         P12& 26&  Sound Engineer, Subject Teacher& Apple Health, Flo, Womanlog& Regular& Yes\\
         P13& 30&  Digital Analytics& Natural Cycles & Absent& No\\
         P14& 25&  AI and Human-Centered AI& Flo & Regular& Yes\\ \bottomrule
    \end{tabular}
    \Description[Demographic table for 14 interview participants, showing age, educational background, menstrual tracking apps used, period type, and current MCTA usage.]{Table of demographic data for 14 interview participants (P1–P14). Columns include: participant number; age (all between 25–34); educational background (fields such as Visual Arts, Cognitive Science, Medicine, Subject Teacher, Economics, and AI); menstrual tracking apps used (Clue, Flo, Stardust, Natural Cycles, Maya, Apple Health, Huawei Health, Menskoll, and Hormona); period type (Regular, Irregular, or Absent); and current MCTA usage (Yes or No).}
    \caption{A table showing participant demographics for 14 interview participants (P1-P14), with columns for participant number, age (ranging from 25-34), educational background (including fields like Visual Arts, Cognitive Science, Medicine, Subject Teacher, Economics, and AI), the menstrual tracking app(s) they have experience using (including \textit{Clue}, \textit{Flo}, \textit{Stardust}, \textit{Natural Cycles}, \textit{Maya}, \textit{Apple Health}, \textit{Huawei Health}, \textit{Menskoll}, and \textit{Hormona}), Period Type (Regular, Irregular, Absent), and if they are current users of MCTAs (Yes or No).}
    \label{tab:my_label}
\end{table*}

We recruited 14 females aged 18-35 (average = 28) for the interviews, who were current or former MCTA users and who had experience with AI features in MCTAs, regardless of period type (regular, irregular, absent). We excluded MCTA users with no prior experience with AI features. The age group was selected due to the high level of digital literacy in this demographic \cite{eurostat2023skills}. Digital literacy was appraised by participants' self-reported knowledge of MCTAs and AI. Recruitment occurred through the first author's personal networks and through an advertisement posted to closed, female-oriented Facebook groups for young females. The study followed the ethical protocol of the hosting university of the first author. The interviews had a mean duration of 24 minutes. 

\subsubsection{Procedure}
\begin{figure*}
    \centering
    \includegraphics[width=0.9\linewidth]{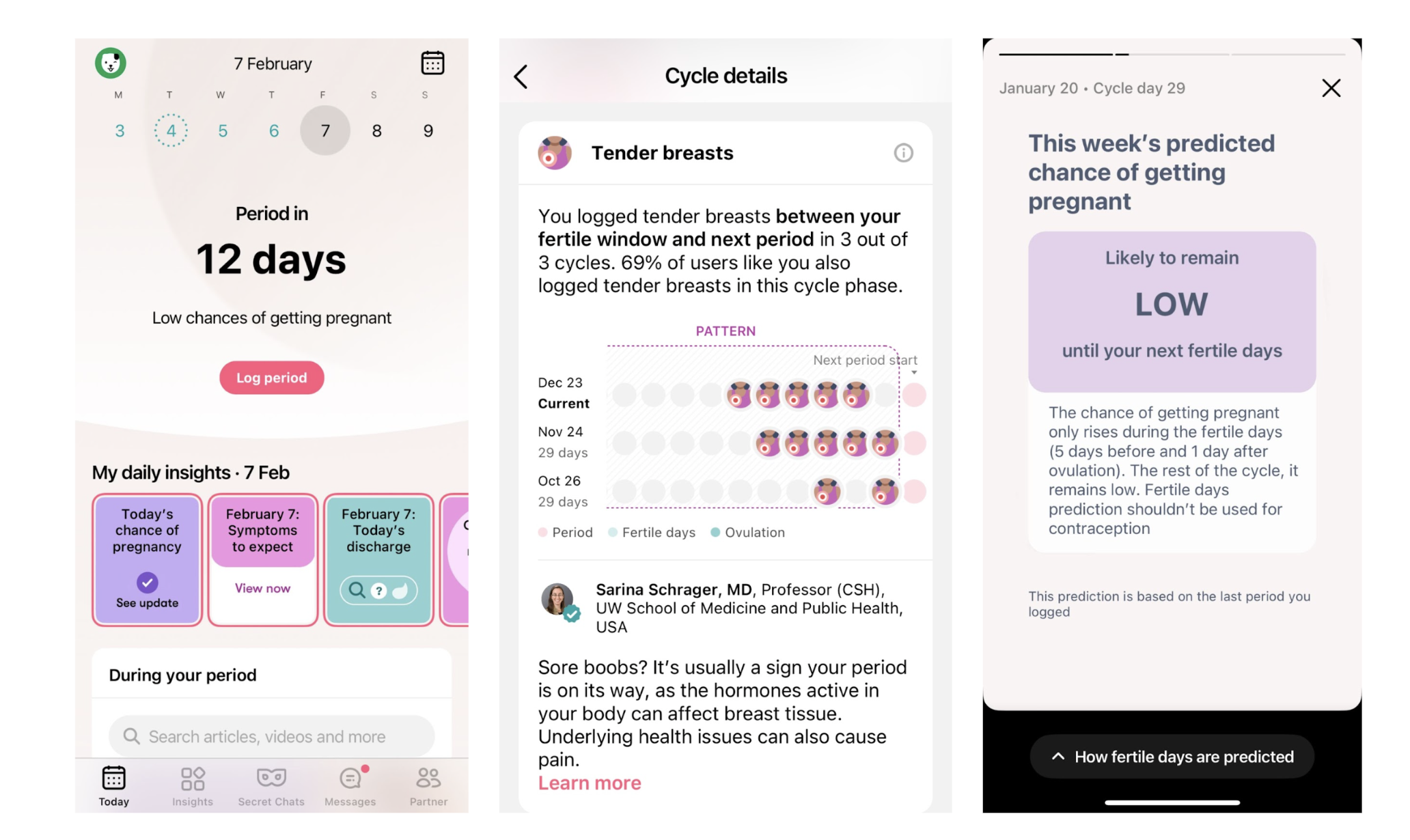}
    \Description[Three screenshots from the Flo app shown to interview participants, illustrating AI cycle predictions, symptom insights, and fertility prediction features.]{Three smartphone screenshots from the Flo app were shown to the interview participants. From left to right: AI-based cycle and period predictions displayed in a calendar view; personal symptom insights and predictions with icons and descriptions; and fertility prediction features with explanatory text.}
    \caption{Three smartphone screenshots from the \textit{Flo} app shown to interview participants, displaying: AI-based cycle and period predictions with calendar views, personal symptom insights and predictions with icons and descriptions, and fertility prediction features with explanatory text.}
    \label{fig:Flo_screenshots}
\end{figure*}
Before the online interviews, participants signed an informed consent form. The semi-structured interviews consisted of questions about participants' demographic backgrounds, menstrual cycle-tracking habits, motivation for using MCTAs, and past experiences with AI features and explanations in MCTAs. The interview was further guided by questions designed to examine the participants' understanding, awareness, and lived experiences with AI features and explanations, with a particular focus on mood- and symptom-based AI predictions, including how their logged data influences AI-based predictions and insights. To gain an understanding of how users interpret AI explanations in the interface design, participants were shown three different screenshots from the {\em Flo} app. These screenshots show some of the most popular AI features and explanations in MCTAs (see Figure \ref{fig:Flo_screenshots}), including cycle and period prediction, symptom insights and predictions, and fertility prediction. The {\em Flo} app was chosen for extracting screenshots to represent AI features since it is one of the most popular MCTAs, and thus most likely to be familiar to interview participants. The purpose of showing the screenshots was to examine how AI explanations are understood, while the more open interview questions were aimed to examine users' lived experiences. The interviews were conducted in the participants' native language Swedish or in English, at the participants' preference. 

\subsubsection{Data processing and thematic analysis}\label{interview_analysis}
All interview recordings were manually transcribed by the first author and translated into English, and identifiable information was anonymized. The analysis of the interview transcripts was guided by Braun and Clarke's  \cite{BraunClarke2006} six-phase framework for reflexive thematic analysis. Reflexive thematic analysis allows researchers to consider the implicit and underlying meanings in participants' reflections while recognizing the inherent subjective lens, which aligns with phenomenological approaches. The coding process employed an iterative approach, involving multiple rounds of collaborative coding and organizing code into groups. Before the group autoethnography, Author 1 developed an initial codebook by coding interesting topics and quotes across all interviews to gain familiarity. Authors 1 and 2 then collaboratively developed a new codebook based on their understanding of the data regarding entanglement in MCTAs. Both authors jointly reviewed 20 percent of the interview data to establish a shared understanding of the codes and code groups, and recoded all interviews. The codes were analyzed collaboratively in relation to the research question to identify narratives and themes. The themes developed were discussed among all four authors to establish a shared understanding of the data before commencing the group autoethnographic study. The codes, code groups, and themes were further developed after the start of the autoethnographic study based on the author's evolving and deepening understanding of human-AI entanglement in MCTAs. Direct quotes were modified by removing filler words and repetitions.

\subsection{Results}
To identify the types of understandings and experiences that emerge as users engage with predictive AI insights and explanations regarding moods and symptoms in MCTAs, we analyzed data from 14 semi-structured user interviews. Two main themes emerged: (1) Users have an understanding of AI features and Data Collection, (2) Users' lived experience is entangled with predictive AI features.

The first theme is related to how users understand predictive AI features and data, with the sub-themes: (1.a) Intuitive grasp of AI features despite limited explanations, and (1.b) Awareness of the effects of individual and collective data on AI predictions. It provides the context for the second theme. The second theme is related to entanglements with predictive AI features, with the sub-themes: (2.a) Co-constructing lived experiences with AI predictions, (2.b)  Validation of negative lived experiences from AI predictions, and (2.c) Negotiating mismatches.

\subsubsection{Theme 1:  Users have an intuitive understanding of AI features and Data Collection. \\} \label{theme1}

\textbf{\textit{Intuitive grasp of AI predictions despite limited explanations.}} Broadly, participants demonstrated an intuitive grasp of predictive AI features in MCTAs, though this understanding was frequently tempered by doubt and lack of engagement. Since MCTAs generally avoid using explicit AI terminology in the UI and marketing materials, many participants relied on guesswork to distinguish AI-mediated features from standard system functions. 

For most participants, their intuition is correct. For example, participants stated \textit{“I’m guessing the predictions on when I will get my periods [is AI]” (P1), “push notifications about all sorts of things related to your cycle, I think that's AI-based advertising” (P8), }and\textit{ “[logging] choices for both symptoms and mood, if you click on them, [the app] after logging for a while, can produce predictions, and I think that counts as AI, or is that not AI?” (P10).} Participants' intuitive understandings were derived from multiple sources. When asked how they identified AI-based features, they referenced app onboarding phases or AI explanations (P7), prior education in AI (P10), and prior work experience as a digital analyst on AI-based self-tracking applications (P13).  

Participants (P1, P3, P9, P11, P13, and P14) understood the limitations of predictive AI insights (e.g., the assumption of cycle regularity), even though the apps themselves did not provide such information. For example, P1 and P11 were aware that period predictions may be inaccurate for irregular cycles, as P1 explained, \textit{“I don't have regular cycles so I’m always aware that it might change. So, I know that if [the app] says that I might get my period, I know it might not be correct, but it gives me a good indication”.}

Furthermore, some participants (P1, P3, P9, P13, and P14) correctly identified that the accuracy of AI predictions depends on the user’s logging practices, referring not only to menstruation tracking but also to mood and other symptom tracking. For example, P3 explained, \textit{“I assume that if I don't enter things every month when I get my period, then it will know less or make a worse prediction”}. P9 shared a similar thought: \textit{"The more data you log, the more accurate it becomes (...) or the more it gets to know you”}. In this sense, users have an intuitive understanding of how AI-based predictions work and how the system learns and improves over time. 

Many participants (P2, P4, P8, P9, P10, P11, P12) had either tried multiple MCTAs before, or were currently engaged with multiple ones simultaneously. One motivation for using different MCTAs was to compare their accuracy, as P2 shared when asked; \textit{"I haven't been able to rely on one [app], so I've wanted to have several at the same time to see if they match or differ on [predictions on] when your period is due and your mood (...) My period is a bit irregular now, it has been more regular before. It means that the apps don't really keep up or know when you're going to get your period, so, I check all of them and find the average.}". This demonstrates an awareness that predictions can vary between MCTA.

Although most of the participants appeared to have an intuitive understanding of which features were AI-based and that predictive AI insights are related to their logging, they expressed explicit uncertainty about AI features in general (P2, P4, P11, P13, and P14) and how they actually worked (P2, P3, P4, P9, and P14). For example, when asked about how users' data affects predictions, P13 stated, \textit{“As I said, that information is not evident. I might be concluding things that isn't correct (...), you can only assume”.}  Similarly, P9 shared\textit{ “Now when I look at it and think about it, I don't know (...) I've used this app for several years and I don't know how much I trust it”}. One participant even expressed a complete lack of awareness --- \textit{“I didn't even know that these apps I use contain AI”} (P11). When asked if they feel that they understand how predictions work or how their logged data affects it, P14 responded, \textit{“I guess I have some sort of basic understanding of it, but definitely not in detail, definitely not a deep understanding, no”}. P4 expressed not understanding how their data or actions affect the predictions, stating,\textit{ “No. I can't, I would be lying if I said otherwise. No, not really aware at all.”}.

In addition to the apps' opacity regarding predictive AI, some participants (P4, P9, P10, and P11) expressed notable ambivalence or a general disinterest in how these features work. As P9 put it, \textit{“they haven't explicitly told you, or I haven't bothered to find out or read about it either”}. P3 further elaborated on the reason, \textit{“no one bothers to sit and read through thirty pages of terms and conditions”}. P11 also indicated this lack of engagement with AI explanations, \textit{“maybe they have some kind of quick guide on how these features work if you want to use the app for this and that, but it's nothing I've put any effort into checking”.} As P10 summarized, \textit{“since I think it's correct so often, I just trust that the technology works, whether it's AI or if it's fake AI. If it's completely made up... Well, it works for me anyway”}.

In summary, participants demonstrate an intuitive understanding of how AI predictions work and how their logging practices and data can affect prediction accuracy. Although the participants' understandings were accurate, they lacked certainty and specificity. When probed, they were unable to specify their understandings or confirm their assumptions due to a lack of AI explanations and transparency in MCTAs. This intuitive understanding, developed without explicit AI explanations, may cause users to more readily trust AI predictions and seek validation of their lived experiences (\ref{theme2}).

\textit{\textbf{Awareness of the effects of individual and collective data on AI predictions.}} Participants (P1, P2, P8, P9, P10, P13, and P14) were aware of the impact of other users' data on personalized menstruation and symptom/mood predictions they received in the apps. This awareness originates from AI explanations of the app and through a knowledge of how machine learning models are trained. For example, P2 recalled from memory, “\textit{The apps I have used can add ‘This is how others can feel during this period’.”} and P8 examined their app during the interview and reasoned \textit{“So [the prediction] is based on others' [data] that are now passed on to me for a supposed recognition factor I think”.} P1 explicitly expressed a liking for seeing other users' data: \textit{“I like the feature where it brings in the other users”}. 

Participants (P9, P10, and P13, P14) reasoned in the interview about the presence of predictive AI features and the impact of other users' data, drawing on their knowledge of how machine learning algorithms generally work. For example, P10 explained, \textit{“probably some kind of average somewhere (…) [the app] have some basis to go on and then I somehow feed it so that it learns more about how I work. But that's just guessing”}, and P14 guessed, \textit{“...that the system is based on other users' data, but they don’t clarify if that data influences the output or not”}. 

Participants' awareness of how their personal data might be used in MCTAs affected their attitudes and reasoning about data privacy, leading to selective data logging. P5 shared that they avoid using AI-based features; \textit{“One reason I didn't want to use the more advanced features is that I don't want them to have it [personal data], to resell information on how I'm feeling or when I'm craving sweets, or all these things (...) that's probably why I chose not to use those other features”}. P14 explained this concern thoroughly: 

\begin{quotation}
\textit{"I think one huge issue is that [the data logged] is very sensitive data. It’s a lot about your health, both physical and mental health. And I know that you can also input data about your sex life, and that’s probably also very, very sensitive data, so… I really like the idea, but I think data privacy is a huge, huge issue with these types of apps. And I feel that’s often a topic that's not really talked about enough. I can only speak of Flo, but they of course, always promise you that they won’t ever give your data to anyone. But Flo has a free version, and you don’t see any ads, so the question always is: How do they make their money? (...) They likely use your data. So I definitely think about what data I want to log and what I don’t want to log. And I think if it were only about knowing more about my period, I would definitely input everything. But, knowing that it is a company behind, and I’m not really aware of what kind of data they will forward - I’m not logging everything that I could log."}
\end{quotation} 

Participants' (P9, P5, and P4) awareness extended to the potential of their data to contribute to research on women's health. P9 explained that even though they were skeptical about how data might be handled, shared, or sold by MCTA companies, they hope that their data will contribute to research on women's health: \textit{"It would be good if you could learn more about what your data actually contributes to, because you can hope that some retired doctors want to understand the female body better (...) I use this AI app that I am fundamentally against, but I am actually contributing to something good"}.

Thus, participants demonstrated awareness that their data is part of a collective dataset used for machine learning predictions in MCTAs. They are further skeptical about how their personal data might be shared or sold by MCTA companies, but hope that their data contributes to research on women's health. 

\subsubsection{Theme 2: Users’ lived experience is entangled with predictive AI features. \\} \label{theme2}

\textit{\textbf{Co-constructing lived experiences with AI predictions.}} By living with and regularly interacting with MCTAs, users benefit from increased self-knowledge and awareness of their menstrual cycle and various PMS experiences, while also becoming interconnected with predictive AI features. The interconnectedness blurs the boundaries between the app's AI predictions and users' lived experiences.

Participants (P2, P3, P4, P8, P11, P13, and P14) described being motivated by increased self-awareness of their menstrual cycle through living with the technology, and expressed appreciation for how these apps support cycle awareness and predict when to expect their period or various PMS experiences. 
Even as their cycles change over time, such as with the use of birth control pills or experiencing irregular cycles, the interview participants continue to find educational value in tracking and personalized predictions. 

Compared to more tangible physical symptoms, participants reported a particularly complex relation between AI predictions and their lived experience around mood and mental states. P2 and P14 described how the MCTA contributes to increased awareness of their felt experiences, in particular, of cycle-related mood changes. As P14 shared, \textit{“I would say for physical symptoms, I’ve first been aware of them, because if you’re in pain, you are in pain. But for mental symptoms, I think the app has made me more aware of them”.} The MCTA and technology-derived knowledge became integrated into their lived experience as a tool for understanding their bodily sensations and emotional states in relation to their menstrual cycle. For instance, P14 explains, \textit{“I got more aware of possible symptoms or mood swings through the app”}, which they also believe has had a long-term effect; \textit{“By now I’m usually more aware of myself”. }

More notably, multiple participants reflected on the interconnectedness between receiving predictions about their mood and symptoms and their subsequent felt experiences. Those predictions could be personal predictions or general information about how most users feel at this stage in the cycle. Participants described that AI predictions feel like an“oracle” or a “self-fulfilling prophecy,” affecting how they perceive their mood. P9 expresses, “\textit{it becomes like a horoscope, a bit of a self-fulfilling prophecy”}, and continues to explain that it can lead to negative spiraling --- \textit{”I can also get a little stressed out from [becoming aware of] where I am in the cycle, because you see ’Now it's ovulation’, and you know that now it's just downhill from here. It's just going to get worse and worse for two weeks. It can psyche me out too.”} Interestingly, P5 does not want to use predictive AI features in MCTAs because of her experience manually tracking cycle-related symptoms with pen and paper. She explained, \textit{“I have previously tried to log my period and how I feel, not via apps, but by writing and reporting it myself, but I felt it became a self-fulfilling prophecy. (...) 'Now I'm in this period of my cycle, so I should feel like this or this', but perhaps I don't really feel like that"}. She concludes that her reason for not using predictive AI features in MCTAs is that \textit{“It might not be great for me to [receive the prediction] 'You usually feel bad during this part of your cycle'"}. This experience indicates that internal and experiential factors highly influence experiences of PMS. With MCTAs' AI features, however, users encounter predictions about their experience more regularly.

However, the participants' reliance on AI is not total or uncritical. P8, P9, P10, and P14 expressed awareness of the distinction between AI predictions and insights and their own self-knowledge, recognizing that multiple factors beyond the menstrual cycle could influence how they feel and how they should act. When P10 was asked about how they felt when AI mood predictions were incorrect, they shared, \textit{"You become thoughtful (...) What is it that affects my mood if it's not my own body? But I don't usually get too caught up in it either; you can have low periods anyway, it could be the weather, it could be that I ate poorly, it could be that I'm hungover for the third day in a row. There's usually some other explanation. But you miss the feeling of [getting] confirmation from the app"}. 

Furthermore, participants express resistance to mindlessly following predictions, asserting their desire to maintain autonomy and make their own decisions regardless of what the app recommends. As P10 expresses around receiving a notification advising them to eat chocolate and chill,\textit{ "I don't need to read about what I should do today, I'm still going to do what I feel like doing. So those things [notifications] are nice to know what you 'should' do today, but I'll still do what I want."} Notably, P5 and P8 explicitly expressed concerns about becoming overly dependent on the app for particular mood predictions, preferring to maintain their own sense of lived experiences, echoing P10's sentiment that they don't need an app to tell them how they feel.


\textbf{\textit{Validation of negative lived experiences from AI predictions.}} Participants (P2, P9, P10, P14) expressed that a key value of receiving AI predictions is that they confirm users' negative lived experiences. As P2 puts it, \textit{“I feel that it's very helpful when it matches with how I feel and what I'm experiencing”}. These experiences are often connected to negative moods as P9 shared, \textit{”If I feel crappy, I usually check where I’m at in the cycle, before I jump [from a height metaphorically ...] it's like "ah but okay, it's not really weird that I hate myself”}. P10 emphasized this point with \textit{”it's actually mostly to get a confirmation or indication of ’Oh I'm sad now, and I should be, because it says here that I'm due for my period in a week’. Specifically, when I'm sad or feeling bad, I want some kind of confirmation that it's okay or that it's not that strange. That you can somehow verify, or justify how you're feeling”}. P14 expressed a similar perspective --- \textit{“... it's usually good for me to be aware of where in the cycle I am, so I know why I might react more tense to stuff I might usually be more relaxed about”}. The desire to validate the lived experiences contributes to the co-construction of meaning through users' interactions with MCTA predictions.


\textbf{\textit{Negotiating mismatches.}} When AI predictions do not align with users' lived experiences, a different dynamic emerges, one that prompts self-reflection and various strategies for managing the mismatch. 
Participants expressed that when they receive AI mood and symptom predictions that do not align with their lived experiences, they engage in self-reflection about the discrepancy or try to manage the predictions. For instance, P10 shares their experience of this mismatch: \textit{”Sometimes you can go in and feel sluggish and tired and weird and then ‘Wow! It says I'm going to ovulate. Why am I not happy and have a God Complex?’ (...) Then I become more thoughtful 'Well, why am I sad now? Am I in some fucking depressive phase or something?' What's affecting my mood if it's not my own body?"} P8 also reflects on this, sharing that they would become \textit{”almost a little disturbed by this app. That according to algorithms, you should feel like this". }

Some participants actively tinker with their data to reduce mismatches. Due to the limited ability for users in MCTAs to give feedback to the underlying machine learning models, P10 manages the mismatch by altering past logged data to align the predictions with their experiences, explaining that \textit{“I usually go and correct [past period dates] so that [the predictions] matches the mood better. Even though I might know that I bled for a long time, I'm like 'maybe it was spotting'  (...) and then [the prediction] actually corrects itself (...) So I think you can correct certain things”}. This behavior indicates a need for users to feel aligned with the AI predictions in their MCTA. Notably, only P2 and P14 expressed doubts about the accuracy of the prediction, rather than tending to self-reflection.


In summary, the results of the interview study revealed that users have an intuitive, rather than explicitly informed, understanding of the AI features in MCTA apps. Through their interactions, they derive conclusions about the role of their logging practices, their own and other users' data, and prediction algorithms. We identified a lack of AI explanations to confirm users' assumptions, leading to limited transparency about which features are AI-based, their limitations, and their capabilities. With this baseline, we identified various enactments of human-AI entanglement. Our data showed that users are inclined to validate their lived experiences through AI predictions, and mostly question themselves rather than the accuracy of AI systems when mismatches occur, which shows a tendency to over-rely on the AI predictions. Most importantly, users evaluate their bodily experiences in light of the predictions they receive, a phenomenon they refer to as a self-fulfilling prophecy. Here, it is difficult to assess whether those experiences are illusions arising from the predictions or objective bodily experiences. Therefore, users' self-knowledge is inseparable from the interactions with the MCTA app. In other words, it is co-constructed \cite{homewood_putting_2020}. The results further highlight the importance given to the alignment with and validation from the apps' predictions and reflective moments created in the mismatches between actual and predicted symptoms. These meaning-making mechanisms go beyond efficiency and accuracy of the apps. The presence of predictive AI features in MCTAs influence bodily sensations, emotions and self-knowledge, making these apps entangled with users' lived experiences. 

Although the interview study provides rich accounts of human-AI entanglement, they are based on retrospective accounts from users with limited awareness of AI features and explanations, and some do not regularly log their mood in their self-tracking practice. To enrich our understanding of the human-AI entanglement through an in-depth analysis of lived experiences, including logging moods and the design of AI interface elements, we conducted a group autoethnography. Further, our expertise in HCI and interaction design allowed a critical and more attentive engagement with these technologies in general. We aimed to collect momentary reflections of our experiences with MCTAs, while accounting for our non-normative cycles and bodies. We explain Study 2 in the next section, and discuss the findings of both studies in Section \ref{discussion}.

\section{Study 2: Group Autoethnography}

\subsection{Methods}
We conducted a group autoethnography as the authors of this paper to enrich our understanding of the human-AI entanglement in MCTAs, with a focus on mood and symptom tracking and living with AI predictions and explanations. Autoethnography captures researchers' subjective and lived experiences combined with their expertise, enabling nuanced reflections  \cite{kaltenhauser_playing_2024, williams_anxious_2015, desjardins_introduction_2021}. Autoethnographic research can also be done by multiple researchers, such as in duos \cite{cecchinato_rethinking_2025}, trios \cite{howell_cracks_2021}, or bigger groups \cite{chen_towards_2024, mack_mixed_2021, cavdir_sonic_2025}. Group autoethnographies enrich perspectives by incorporating multiple positionalities and reducing individual biases  \cite{cavdir_sonic_2025, chen_towards_2024, mack_mixed_2021}. Following this tradition, we followed a group autoethnography process as four authors of this paper. As researchers working at the intersection of self-tracking, AI, and well-being, we find an autoethnographic study fitting to uncover further the phenomenon of entanglement and the processes that lead to it. Furthermore, we represent diverse user groups in terms of menstrual cycles, including regular cycles and irregular cycles with PCOS, and those in perimenopause and menopause, enriching the perspectives in our data.

\subsubsection{Ethnographic backgrounds} 
    Author 1 has regular periods. She has been using the \textit{Flo} app consistently for the past seven years, with a few episodic pauses. She tracks the start of her period, as well as any moods and symptoms that are believed to be cycle-related. She experiences regular cycles with predictable patterns, journals daily, and uses MCTAs to understand cycle-related mood and symptoms. She has previous technical education and professional experience in digital product design.
    
    Author 2 has Polycystic Ovarian Syndrome (PCOS, a common hormonal disorder that affects menstruating women \cite{world_health_organization_polycystic_2025}). This syndrome has very individual manifestations, and the most prominent symptoms for Author 2 are irregular periods and skin problems. Her involvement in the project coincided with a period of her life where she was committed to managing her PCOS. She had been experimenting with various menstrual tracking apps for the past two months, and has been using multiple trackers to track the length of menstruation, mental and physical symptoms, consistently but irregularly. 
    
    Author 3 is in post-menopause. She has been using hormone replacement therapy (HRT) for a few years. Her current treatment follows a highly regular schedule that induces a predictable, artificial bi-monthly bleeding. Because her hormonal levels have stabilized, she does not currently feel the need to track menstrual or menopausal symptoms in any systematic way. Before menopause, she used the \textit{Garmin} app for several years to manually track her menstrual cycle, primarily to predict upcoming periods. She had not previously used a dedicated menstrual or menopause tracking app. Her background includes long-term self-tracking practices for other aspects of health, but beginning a new system from scratch for menopause felt disconnected from her existing data, systems, and routines.
    
    Author 4 is in perimenopause at the beginning of the study and has not had menstruation during the study. During the decades before perimenopause, she used several apps, including \textit{Excel} spreadsheets, \textit{Fitbit}, and \textit{Apple}'s Health app, to track her cycle regularly and manage reproductive health. Since perimenopause, she paused the use of MCTAs due to their inaccurate predictions and the anxiety of missed predictions. She chose {\em Clue} for the study because it included a perimenopause mode. Compared to other authors, Author 4 also received AI/ML technical training in addition to a background in HCI and Design.

\subsubsection{Procedure}
Our process started with author 1 drafting a study plan, which was discussed in two iterations among all authors and agreed upon. Then, a one-week pilot study commenced by Authors 1 and 2, who began collecting data through journaling and artifact analysis, following the method planned for the full study. The aim of the pilot study was to determine the feasibility of the data collection method. After the one-week pilot study, all authors collectively discussed the method plan and concluded that it was feasible for the full study. Directly following this discussion, the full study commenced for all authors and lasted 8 weeks. Authors 1 and 2 were already using MCTAs, and the other two authors used apps specifically for the study. 

We conducted check-ins with all authors every other week. During check-ins, we focused on discussing the autoethnographic method and data collection, including clarifying the types of subjective reflections to include in the data collection. This helped align our reflections with the study aim. During check-ins, we avoided discussing personal reflections to minimize the influence on each other's reflections. After the data collection period, we conducted several meetings to discuss, share, and interpret the data collaboratively. We present additional details on data collection and analysis below.

We collected data through two primary methods: journaling and artifact analysis. We kept personal records of detailed reflections and observations of first-person experiences, such as thoughts, feelings, and reasoning processes during app interactions, as well as self-reflective accounts of experiences and perceptions. Further, we captured screenshots of the app's interface and design features, including AI predictions, insights, explanations, and other relevant features corresponding to journal content when appropriate. The journaling process was guided by three questions designed to avoid priming effects: (1) "How do I feel today?", (2) "What did I do?", and (3) "What did I experience?".

We designed the autoethnographic study to reflect authentic app use, meaning that we did not restrict or demand engagement with specific AI features, but allowed each author to engage with MCTAs as they would in an everyday contexts, to identify organically emerging experiences. Due to our diverse cycles, we wanted to allow each individual to relate to MCTAs based on their own needs. We conducted the study using four different MCTAs based on our individual needs and preferences, as each MCTA markets its product to different user groups. Author 1 continued to use {\em Flo} as a long-term user; Author 2 used \textit{Clue} and \textit{Stardust}; Author 3 used \textit{Balance}; and Author 4 used \textit{Clue}. All four apps that we used were assumed to incorporate AI-based features such as predictions and insights. However, the precise nature of how machine learning or data is used in these applications remained uncertain as their companies did not reveal details of their proprietary models. Therefore,  we operationalize the term "AI features" based on our subjective interpretations in this study.



\subsubsection{Data Analysis}
Each researcher conducted a first round of data analysis to identify themes from their own autoethnographic reflections, given the intimate nature of the collected data, and to ensure each author's privacy and the right to review, control, and exclude any sensitive or personal information. The individual findings were then collaboratively discussed and shared verbally or in writing with the first author, in the form of themes. The first author synthesized the data while considering ethnographic backgrounds such as cycle type (regular, irregular, perimenopausal, menopausal) and earlier experiences with AI features in MCTAs. This led to an initial set of themes derived from the autoethnographic study. The insights were discussed and reflected on among all authors to ensure the correct interpretation of subjective experiences. By the start of the group autoethnography, the major themes from the user interviews had already been identified, so the insights from the group autoethnography were reviewed against earlier themes. New findings from the group autoethnography that had not been identified in the previous interview study were presented as new results, while common findings from both studies were integrated into the discussion section to formulate design implications.

\subsubsection{Ethical Considerations and Positionality Statement}
Given the highly personal and sensitive nature of the study subject, we paid particular attention to anonymity and confidentiality throughout the group autoethnographic process. The study implemented a self-directed approach to inclusion. Each researcher's data remained private before collaborative discussions, allowing individual researchers to review, protect, and exclude sensitive or personal information before sharing it with other members of the research team. To ensure transparency, we describe our backgrounds in the section above. Finally, we remained attentive to how our positionalities shaped our reflections. As HCI and design researchers at Nordic universities, we brought design and ML expertise beyond that of typical users. Additionally, having conducted Study 1 (user interviews) before this autoethnography, we had already identified a process of entanglement, which risked confirmation bias in our observations. We mitigated this by focusing on the autoethnography's distinct goal: to examine the first-person, in-depth experiences of entanglement rather than simply confirming its existence. We documented our experiences and avoided collaborative discussions of individual experiences before the analysis stage to prevent biasing each other's reflections. However, we acknowledge that our dual role as researcher-participants inevitably shaped what we attended to and how we interpreted our experiences.


\subsection{Results}
The autoethnographic study reveals the first-person experiences of diverse users with regular cycles, irregular cycles, perimenopause, menopause, and hormone-treated bodies. Three novel themes emerged from the group autoethnographic study: (1) Logging as an Entangled Experience, (2) Living with the AI Predictions, and (3) Leaving out the Non-normative Body.


\subsubsection{Logging as an Entangled Experience.\\}

As part of our autoethnographic inquiry, we used the mood-tracking features and logged other felt and bodily symptoms. We observed the algorithmic logics of MCTA apps, which require categorization, simplification, and predictability, and influence overall interaction dynamics, including logging.  Further, we immediately realized that the apps we used approached the notion of "mood" differently, offering different selections and interaction mechanisms. For instance, \textit{Clue} featured facial expressions, and \textit{Stardust} allowed for logging of intensity. 
A1, A2, and A3 observed that the interface design and the language used to label moods and symptoms affected their experiences in various ways. A3 felt that the app's categorical structures for logging felt conceptually "weird" for her as they misaligned with how she understands her emotional experiences. More specifically, the app proposed "anxiety" as a separate category from mood tracking, where, for her, feeling anxious was part of "moods". A1 and A2 noted that the language and icons used to label moods and symptoms in the logging interface guided, framed, and shaped how they made sense of their lived experiences. For example, A2 was affected by the icon designs in the logging interface, noticing that she matched her feelings with the icons' facial expressions, rather than the textual labels, which influenced her choice of logging. Similarly, A1 experienced anxiety and planned to log it as such in the MCTA. However, in the app, she received a suggested log based on a prediction of "obsessive thoughts". She felt this label was close enough and thus logged "obsessive thoughts" instead. This change of labeling was preferable as it framed her experience less negatively, since she felt that "obsessive thoughts" was a more neutral and clinical label for her experience than "anxiety". This was also an instance where receiving predictions immediately shaped the logging experience, creating a feedback cycle prone to error. 

The act of logging was also affected by motivation. For instance, A1 noticed that she tended to log only moods and symptoms in the MCTA that she suspected were due to her menstrual cycle (e.g., mood swings, depression, anxiety, or fatigue). A2 had a slightly different experience, where she was exploring the purpose of logging moods. It was towards the end of the study, where she finally concluded that for her, it was only meaningful to log moods that she suspected was related to her menstrual cycle; those are unexplainable through her daily life events. She noted the apps' interfaces does not guide the mood tracking process, and the logging activity was prone to reporting of moods that do not encapsulate the overall experience of a given day. A4 reflected that her way of tracking mood was one-day retrospective, which allowed her to analyze her day to arrive at meaningful, reliable insights about the overall mode of the previous day, a conscious tracking activity that the apps' do not support.

We explored logging as a critical part of interacting with MCTAs and predictive AI, an aspect that did not fully surface in Study 1. Our autoethnographic accounts revealed that logging is far from an objective reporting of our lived experiences and is already shaped by our motivations, the interface design language, and the predictions we receive. This raises a critical concern: the predictions we receive are calculated through inherently faulty logging practices for the hard-to-categorize nature of subjective experiences, such as mental symptoms and moods. These predictions have the potential to shape our embodied experiences, as we reflect on it more below.



\subsubsection{Living with the AI Predictions.\\}
During the autoethnographic study, three of us (A1, A2, and A4) experienced instances in which, after receiving AI predictions of specific symptoms from the app, we subsequently experienced those symptoms. For example, A1 noted multiple instances of receiving predictions of headaches that she experienced immediately after using the app or later during the day. These experiences increased her trust in AI predictions, prompting her to reflect on whether her highly regular cycle, combined with her 7-year use of the \textit{Flo} app, has enabled the ML model to know more about her body than she knew herself. Reflecting on the findings from Study 1, she then attempted to learn to distinguish between felt experiences that may have been influenced by the app's predictions and those that would have occurred without AI influence. However, she found it was impossible to be sure whether it was a placebo effect, and instead felt that the app's prediction might be incredibly accurate. A4 noted receiving a prediction for cramps, which she experienced about 1 hour later. Her interpretation of this event was that the app made them more aware of a bodily sensation they might otherwise not have noticed. A2 received a prediction one morning that they would feel great and highly energized throughout the day, which she subsequently experienced. She felt unsure about the accuracy, but thought the prediction acted as a positive mental probe. 

It was interesting for us to observe that, even with our critical approach to these technologies, we were unable to unpack the impact of receiving predictions on our lived experiences. Our experiences were realized only after encountering the predictions, which all three of us interpreted differently. This phenomenon highlights the complex and subjective nature of living with these technologies, where truth is not easily uncovered but rather lies in our interpretations. Taken together with inherently non-neutral logging experiences, our realities are co-constituted through a complex cycle of foundationally limited predictions. Through our first-person experiences, we also validated that human-AI entanglement not only needs to be unpacked, but should be designed for.




\subsubsection{Leaving out the Non-normative Body.\\}

Those of us without a regular cycle, A2, A3, and A4, were motivated to use MCTAs to gain a deeper understanding of ourselves. However, the systems we used failed us in different ways due to their lack of adaptability and predictive features for users with non-normative bodies and cycles. A2, who has an irregular and unpredictable cycle due to PCOS, referred to this as "regularity language", which shapes the logging design, predictions, explanations, notifications, and the app's interface design in general, resulting in a negative experience for her. A2 started the study with a "late period", and the delay was more than a month at one point in the study. For most of the study, she observed that the apps' design languages kept framing her period as "late", even though both apps were used in "irregular period" mode. She was frustrated with the framing of "lateness" as used in opposition to being on time or functioning well. She felt this language framed her irregular body as faulty, or something to be fixed. She noted that the \textit{Stardust} app sometimes framed her cycle in more neutral terms, such as "This is cycle day 92," rather than counting the number of days her cycle differs from a normative cycle, which eventually led her to use this app much more. After her period was more than 30 days "late", the \textit{Stardust} app's design changed, removing the indication regarding the temporality of lateness. While this felt good and more inclusive, she noted that the lack of explanations or acknowledgment of the reasoning behind the interface design change left her feeling confused and empty.

A4, who is in perimenopause, experienced little need to track her cycle using an app, since the app kept predicting her period to always be "just around the corner", supposedly consistently late, although the app was operating in a specially-offered perimenopausal mode. However, while the app offered specific logging options such as "trouble falling asleep" and "hot flashes," she was not sure how the app actually accounted for this transitional stage, which varies significantly between and within individuals. It, in turn, reduced her level of trust in the app's predictions.

A3, in menopause and on hormone replacement therapy, downloaded the \textit{Balance} app for this study, marketed specifically for menopause and hormones. A3 was motivated by the potential to gain insight into the patterns between her mood changes and hormone cycles. However, she was disappointed that the app did not provide any tools for this. She did not receive any AI predictions, explanations, or personalization based on her logging, which made her feel the app did not meaningfully process her data. She reflected on whether she had missed something in the app. She also thought that the period prediction features in the MCTA were irrelevant, as her bleeding schedule is medically induced. Thus, she felt marginalized, and the app offered little personal value. Hence, her logging became something she felt she did out of obligation rather than out of intrinsic motivation.
Across these three experiences, a common thread emerges: MCTAs are designed around an implicit norm of cyclical regularity, and when bodies deviate from this norm, the apps either pathologize the difference, fail to adapt to transitions that are part of the menstrual life course, or provide no value at all. Together with the already entangled meaning-making through these apps' design features, predictions, and explanations in place, this underlying normative assumption can be felt not only as an algorithmic exclusion but as a bodily, lived limitation. Our experiences with these apps provided first-person perspectives on the hidden harms these systems can cause, even through limited interactions.

\section{Discussion} \label{discussion}
In this final part, we discuss our findings in relation to earlier studies and their implications for design, suggesting alternative approaches to designing AI explanations and predictive features in MCTAs that build on a deeper understanding of human-AI entanglement. 


\subsection{Processes of Human-AI Entanglement}

Our results show that users engage with AI predictions as an integrated tool to gain self-knowledge, relate to their cycle-related moods and symptoms, and to validate negative lived experiences. This support earlier findings by Grimme et al. \cite{GrimmeMyData2024} that women track their cycle data to gain self-knowledge and desire reassurance and validation that their perceived cycle-related experiences are accurate. We further identified that the desire for validation could contribute to a tendency to over-rely on AI predictions in mood and symptom tracking with MCTAs, expanding on previous research that found that users doubt, reinterpret, and override their lived experiences in favor of algorithmic predictions \cite{zhu2025centers, obrien_periods_2025}. Towards this end, Fox et al. \cite{fox_multiples_2020}, and Tylstedt et al. \cite{tylstedt_reimagining_2023} present uncertainty as a design approach to support bodily sense-making over prescriptive predictions. We identified existing instances where users make use of mismatches between AI predictions and their lived experiences as self-reflective instances, highlighting these systems' capacity to support self-knowledge even through seemingly failed predictions, pointing towards an alternative design space.

The key finding of our research is that the users' lived experiences can not be studied separately from their interactions with MCTA apps. Our findings explicitly show the phenomena where users evaluate their bodily experiences together with the AI predictions in the MCTAs. Moving beyond a binary understanding of cause and affect, we approach to this phenomena as \textbf{human-AI entanglement.} Multiple interview participants described a “self-fulfilling prophecy”, indicating that receiving AI predictions can influence, affect, or prime users to experience the predicted mood or symptom after using the app, either immediately after or hours later. In Study 2, three out of four authors experienced symptoms, including physical feelings (headaches and cramps) as well as mental states, {\em after} seeing AI predictions. This suggests that even phenomenon with physical sensations can be influenced through entanglement. We further uncovered the negative impacts of binary and normative assumptions of MCTAs, where exclusion of non-normative users was perceived not only as a limitation, but a felt experience in this entangled existence with MCTA technologies and their predictive AI features \cite{sambain2025}. 

The predictions that are the subjects of this entanglement depend on users' data logging. The menstrual cycle alters bodily and hormonal physiology, thereby affecting a myriad of physical and cognitive symptoms that the user can track. Previous research has noted that most symptom-tracking features in MCTAs rely on normative, predefined categories  \cite{lupton2016quantifiedself, tylstedt_reimagining_2023, homewood_putting_2020, reime2023walking}. Our results support previous research, finding that suggestions and predictions for logging, as well as the interface design (including labels, wordings, and icon design), influence and guide users' logging and self-knowledge. We further extend these findings to the context of AI predictions, where the interface design represents the underlying assumptions and structures of ML models. These often subjective and intimate symptoms can vary significantly from person to person, and Lupton argued that quantifying menstrual cycle experiences through these categories can make users' lived experiences rigid and dull \cite{lupton2016quantifiedself}. Tylstedt et al. noted that users' logging experience in MCTAs is rarely subjective beyond free-text input, leaving little room for nuance and complexity \cite{tylstedt_reimagining_2023}. We highlight the impact of these limited logging mechanisms on predictions that directly co-constitute users' lived experiences, creating a complex, potentially faulty vicious cycle of self-fulfilling prophecy. 

The logging data not only influences personal predictions but also those of the broader user base, creating a collective data entanglement (see Section \ref{theme1}). Interview participants showed awareness of this collective entanglement and concern about how their data would be shared or sold, but were motivated to share data with hopes of contributing to research on women's health, in alignment with earlier findings \cite{GrimmeMyData2024, HudigDataSharing2025}. As logging practices are never entirely accurate, predictions informed by potentially faulty collective data then affects users' interpretations of their own bodies \cite{homewood_putting_2020}. 

Our results show a lack of transparency across all aspects of AI deployments in MCTAs, including explanations of which suggestions are AI-generated, the limitations of the predictive models, and the underlying assumptions embedded in them \cite{pichon_messiness_2022}. Although current AI explanations address AI features in general, such as informing users that when they log a symptom, the systems try to find a pattern (See Figure \ref{fig:AI_Features}d), and they are currently insufficient. Thus, our results indicate a deeper misalignment between current explainability approaches and how users are entangled with predictive systems. While we do not approach human-AI entanglement as a negative phenomena to be eliminated, the lack of AI explanations may contribute to over-trusting the systems' capabilities and limit opportunities for users to understand and critically engage with their entanglement with AI predictions. This absence becomes more critical when mismatches occur between lived experiences and AI predictions, as MCTAs do not provide AI explanations or mechanisms for users to give feedback to predictive models and correct inaccurate predictions, supporting the vicious cycle. Further, it also contributes to the experience of exclusion of non-normative menstruating bodies, as the lack of acknowledgment of limitations contribute to the perception of the systems as representing ideals. We present an in-depth understanding of the interfaces and interactions where explanations may be necessary, extending the suggestions for implementing AI explanations to MCTAs \cite{zhu2025centers}.

This study offers novel insights into the processes of entanglement between users' lived experiences and predictive AI, shaped by the lack of AI explanations and the interface design of MCTAs. The insights reveal a complex interplay among limited AI transparency, intuitive user understandings, subjective logging practices, experiences of non-normative bodies, and meaning-making processes deeply affected by mood and symptom AI predictions. These patterns of entanglement, together with validation-seeking and self-reflective behaviors identified in our data, raise essential questions about how users come to understand and trust AI predictions in the absence of clear AI explanations. Below, we present design implications to address the key concerns we identified in this work.

\subsection{Designing for Entanglement}
Our study found that entanglement occurs throughout the user journey, from symptom logging to interpreting predictions, yet current app designs fail to account for this interdependent relation. This design weakness presents an opportunity to redesign MCTAs in ways that acknowledge and work with entanglement. 


First, we need to {\bf design user interaction with predictive AI in ways that support users’ development of self-knowledge}. Our studies reveal scenarios where the interaction design of predictive AI in MCTAs may foster over-reliance. This is especially true during the critical moment of logging, when users are extra vulnerable to having their experiences shaped by algorithmic output. We could replace logging suggestions based on predictions with predefined labels and categorical norms in the same order every time. To further support independent reflection, the main logging interface could be designed to include free-text input, allow users to create custom logging items (by naming their own labels and choosing icons), and encourage users to log their experiences before being exposed to AI predictions. This proposal builds directly on current research on "cognitive forcing" in the domain of AI-assisted decision-making \cite{buccinca2021trust}, to slow down human-AI interaction and promote more deliberate use of AI. Furthermore, building on Homewood et al.'s "removal as a method" \cite{homewood_removal_2020}, we speculate on an MCTA that gradually fades in and out of its AI predictions over time to calibrate users’ reliance level on them. 

Second, we can {\bf design human-centered AI explanations to promote appropriate levels of trust and reliance on  AI predictions}. AI explanations can increase the transparency of when predictions are used and how uncertain they are. For instance, in the logging experience of \textit{Flo}, the app's user interface (See Fig. \ref{fig:AI_Features}a) does not provide any explicit information to the user that suggested loggings are based on AI predictions. We propose that MCTAs add AI explanations to make it transparent to users that the suggested symptoms for logging are based on AI predictions and may be inaccurate. 
In addition, designers can use AI explanations to surface the inherent limitations and uncertainties in AI predictions rather than hide them. For instance, an app could explicitly invite users to provide feedback on the AI predictions and their "accuracy". 
These AI explanations could be progressively disclosed to avoid overwhelming the user as proposed by Muralidhar et al.  \cite{muralidhar2025operationalizing}. 

For users with non-normative cycles, we propose adding personalized AI explanations, including providing the option to opt out of AI predictions, extending proposals by Reime et al. \cite{reime2023walking} that MCTAs should \textit{"seek to design for a plurality of reproductive bodies"}. For example, in the settings section of an app, MCTAs could display an informational text tailored to users with non-normative cycles, explaining that predictive AI features might not be accurate for them and providing users the option to deactivate them. In the long run, the HCI research community should invest in more inclusive models to decenter normative bodies and explore different rhythms and human-AI relations \cite{zhu2025centers}. Our findings provide evidence on how human-AI entanglement in MCTAs perpetuates normative notions of body regularity through predictions and interface design. We invite design researchers to build on this foundation and explore alternatives, such as through decentering as a method. 

Finally, while we can reduce the intended negative influence of AI, humans are inevitably entangled with the technology we regularly use. We urge designers of AI-based well-being technology to {\bf design with {\em a priori} assumptions of human-AI entanglement from the start}. This means that we need to be cautious of binary notions such as "accuracy" and "error”, which appear in design guidelines for human-AI interaction (e.g., \cite{amershi2019guidelines}). As our studies illustrate, when users' self-knowledge is deeply entangled with AI predictions, "inaccurate" predictions may become the reality. These cases require design researchers to complicate what an AI "error" means in well-being contexts and how to handle it. Instead, we can design for alternative relations that the users may have with their MCTA predictions. For example, can we design with the intention of making visible users’ entanglement with the AI and giving them tools to decide how to respond to the “self-fulfilling prophecy” they are already experiencing?

These design approaches move beyond simply providing AI explanations to fundamentally reconsidering how predictions are integrated into the user experience. By designing interfaces that acknowledge the co-constructive relation of human-AI entanglement, we can support users in maintaining greater autonomy over their sense-making while still benefiting from AI insights when explicitly desired.


\subsection{Limitations}
This study has several limitations. All participants resided in Nordic countries (Sweden, Denmark, Iceland) or Germany, representing a culturally homogeneous Northern and Western European sample with shared attitudes toward menstruation, bodies, and health technology. Consequently, we acknowledge this cultural homogeneity as a limitation that may affect the generalizability of our findings to other demographics and cultural contexts. 

Additionally, all participants volunteered for the interview studies and may represent users with particularly positive attitudes toward MCTAs and higher levels of digital and AI literacy. This self-selection bias suggests our findings may not capture the experiences of users who are skeptical of or unfamiliar with these technologies. Further, the autoethnographic study was conducted over eight weeks, which was sufficient to observe entanglement processes and patterns. However, a longer-term study spanning multiple years could reveal additional insights on how entanglement evolves over extended periods of use or through significant life transitions affecting menstrual cycles, such as pregnancy, menopause transitions, or changes in contraceptive use. 


\section{Conclusion \& Future Work}
Through semi-structured interviews and a group autoethnography, we investigated how users' lived experiences are deeply entangled with AI predictions. Our results reveal that users understand their lived experiences in light of AI predictions, although these predictions can be faulty due to imperfect logging practices. This leads to  a phenomenon participants described as "horoscope" or a "self-fulfilling prophecy."
In the MCTAs, the user interface features and AI explanations do not support awareness with this entanglement or critical engagement of meaning-making of one's lived experience with logging and AI predictions. Finally, the non-normative MCTA users report a sense of isolation in this entangled interaction. Based on our findings, we proposed design implications for predictive AI features and explanations.

We urge future research to continue examining the processes of co-construction and entanglement between users and AI systems in health contexts, particularly regarding subjective experiences such as moods and mental states. Researchers should explore alternative designs for predictive AI features that support users' critical engagement with how their lived experiences are co-constructed with AI, while maintaining the benefits these technologies can provide. We particularly call for more inclusive designs of MCTAs and predictive features capable of delivering value for users with non-normative cycles and bodies, including those with irregular cycles, PCOS, perimenopause, and menopause. This includes rethinking categorical limitations in logging interfaces to allow for more subjective, open-ended expressions of lived experience. 

\begin{acks}
This research was partially funded by Danish Novo Nordisk Foundation under Grant Number NNF20OC0066119 and Independent Research Fund Denmark (DFF) under Grant Number 5334-00058B. 
\end{acks}

\bibliographystyle{ACM-Reference-Format}
\bibliography{bibliography}

\clearpage

\appendix
\section{Interview Guide}
\textit{Insights on AI features and explanations in menstrual cycle tracking apps}

\subsection*{Personas to Interview}
Females aged 18--35 who have used menstrual cycle self-tracking apps.

\subsection*{General Demographics}
\begin{itemize}
  \item Gender?
  \item Age?
  \item Lifestyle? Occupation, hobbies, exercise habits?
  \item What is your relation to AI and technology (education or interest)?
\end{itemize}

\subsection*{Menstrual Cycle Self-Tracking Habits}
\begin{itemize}
  \item Which app do you use to track your period?
  \item How are you tracking? How often, consistently, or less so?
  \item What types of things do you track? Examples: period-related symptoms, moods, or physical experiences?
\end{itemize}

\subsection*{User Motivation}
\begin{itemize}
  \item What is the main reason why you use menstrual cycle-tracking apps?
\end{itemize}

\subsection*{Experiences with In-App AI Features \& AI Explanations}
\begin{itemize}
  \item What has your experience been with AI features in menstrual cycle-tracking apps? Have you used or encountered any? \textit{Examples: Predictions on when to expect menstruation and symptoms, insights based on your past trackings and other users' collective data, and chatbots.}
  \item How have you found these AI features? Have they been helpful or valuable to you? Why or why not? If not: In what ways?
  \item Have you encountered any explanations of how AI features work in these apps? If yes: Did they help you come to a deeper understanding of the AI features and how to use the app better? If not: Why not?
  \item Do you understand how these AI features work and how your actions affect the output?
\end{itemize}

\subsection*{The Scenario: Screenshots of AI Features from the Flo App}

\textit{``I am going to show screenshots of a menstrual cycle tracking app and some selected AI features. Your task is to just talk aloud about what you are seeing, what you think it is, and your first impressions of it, and I am going to ask you questions.''}

\begin{enumerate}
  \item \textbf{Period Prediction:} When you first open the menstrual cycle tracking app Flo, you see this page.
  \item \textbf{Symptoms Record:} You can scroll down and see these insights on the cycle.
  \item \textbf{Fertility Prediction:} In the app, you can also see other predictions.
\end{enumerate}

\noindent\textit{Ask:}
\begin{itemize}
  \item What do you think this is?
  \item How do you think your own data, if you were a user, would affect this? \textit{[Observe and see if they notice/read/learn from AI explanations]}
  \item \textit{[When/if AI explanations have been identified]} How do you understand these explanations?
\end{itemize}

\end{document}